\documentclass[nojss]{jss}

\usepackage{thumbpdf,lmodern}
\usepackage{eucal, amsfonts, mathtools} 

\usepackage{algorithm}
\usepackage{algpseudocode}

\usepackage{tikz}
\usetikzlibrary{automata, positioning, arrows, shapes}

\usepackage{listings}

\usepackage{framed}

\author{Taylor R. Brown \\University of Virginia}

\title{\pkg{PF}: A \proglang{C++} Library for Fast Particle Filtering}
\Plaintitle{\Plaintitle{PF: A C++ Library for Fast Particle Filtering}}
\Shorttitle{\pkg{PF}: A \proglang{C++} Library for Fast Particle Filtering}

\Abstract{
Particle filters are a class of algorithms that are used for "tracking" or "filtering" in real-time for a wide array of time series models. Despite their comprehensive applicability, particle filters are not always the tool of choice for many practitioners, due to how difficult they are to implement. This short article presents \pkg{PF}, a \proglang{C++} header-only template library that provides fast implementations of many different particle filters. A tutorial along with an extensive fully-worked example is provided.
}

\Plainkeywords{particle filters, state-space models, time series, C++}

\Address{
  Taylor R. Brown\\
  Department of Statistics\\
  University of Virginia\\
  Halsey Hall\\
  PO Box 400135\\
  Charlottesville, VA 22904\\
  E-mail: \email{trb5me@virginia.edu}\\
  URL: \url{http://people.virginia.edu/~trb5me}
}

\begin{document}


\section[intro]{Introduction: Particle Filtering with the \pkg{PF} library} \label{sec:intro}

\subsection{What are particle filters?}

Particle filters are recursive algorithms that are used in the analysis of state-space models, a particular class of time series models that feature latent random variables called "states", "signals", "memory" or "codes."  In many modeling tasks where these types of models are used, the state process represents some intuitive and physically-meaningful process.

Particle filters are used to approximate different sequences of state distributions using weighted samples. Even though they can be difficult to understand and difficult to program, particle filters are often seen as the only option for a large collection of state-space models, and unfortunately, these difficulties has discouraged a wider adoption of their use.

Particle filters work by iterating over observed data in time, and storing a large, constantly-evolving collection of weighted samples. Every moment an observed data vector is obtained, the collection of weights and samples changes. This happens in two steps. First, old particles are "mutated" (or "propagated") into new particles. This means that a user-chosen "proposal" or "instrumental" distribution is used to sample values of the new states using, or conditioning on, the old values. For each new state sample, its corresponding weight is updated. It becomes larger if the sample is "good," and vice versa.

These new samples and weights set the stage for the second step: "resampling." Here, unweighted samples are drawn with replacement from the current collection of weighted samples. Samples with high weights are chosen with a higher probability, and samples with low weights are chosen with a low probability. In this way, samples with large weights tend to end up being duplicated, and samples with relatively small weights tend to be discarded.

\subsection{What does this software have to offer?}

It takes time and effort to implement different particle filters well, and this is true for two reasons. First, the mathematical notation used to describe them can often be intimidating after a first glance. It is easier to obtain a first impression of how these strategies work from a high level view is, but being specific about individual sampled particles requires one to, at the very least, describe which distribution it targets and which index it's being stored at, and so indexes, superscripts, and superscripts with superscripts abound.

Adding to the potential confusion is that there are many variants of the algorithms, and each algorithm is described in different ways by different authors. There are many different notational conventions followed: different letters in different languages represent different distributions and different quantities, some notation is measure-theoretic and some is not, and sometimes the description of the algorithm interchanges the order of the mutation/resampling steps.

The second reason is that, if they are not carefully implemented, particle filters can be quite slow. To become familiar with the algorithms, it is instructive to write a first implementation in whatever language is familiar. However, a first implementation will likely be very slow, and slow particle filters will not be feasible for many data analysis tasks, particularly for ones that involve large, or high-speed data sets.

This paper describes the \pkg{PF} library, the main purpose of which is to make use of {\bf class and function templates} to offer fast implementations for a wide range of particles filters, which can all be used for wide ranges of time series models. Each particle filtering algorithm is provided as a class template with several pure virtual methods. Once a data analyst has a specific state-space model in mind, he will pick which types of particle filtering algorithms to associate with that model. For each model-particle filter pair, he writes a class for his model that inherits from the particle filter's class. Each algorithm requires the ability to evaluate and/or sample from different distributions, and so these are the pure virtual methods that the data analyst must provide.

So, to be able to use \pkg{PF}, one only needs to be able to specify the distributions of his statistical model. The implementational details of each particle filter are abstracted away. Full documentation is available at \href{https://tbrown122387.github.io/pf/}{https://tbrown122387.github.io/pf/}.

This is by no means the first \proglang{C++} library to be offered that provides particle filter implementations. Other options include LibBi \cite{libbi}, Biips \cite{todeschini2014biips}, SMCTC \cite{Johansen:2009:JSSOBK:v30i06} and Nimble \cite{doi:10.1080/10618600.2016.1172487}. The goals of these software packages are different, though--users of these libraries write their models in a scripting language, and that model file gets parsed into \proglang{C++} code.

\pkg{PF}, on the other hand, besides providing a different assortment of particle filters, does not provide a parser, nor does it interface with any interpeted language. It is designed for users that prefer to work in \proglang{C++}. It can be useful for instantiating individual particle filters, but it is also designed in an object-oriented manner to facilitate the implementation of algorithms that update many particle filters in parallel \cite{1453776,1309.2918}. It is header-only, so it is easily added to other \proglang{C++} software projects. The only extra steps are \verb|#include|-ing relevant headers, and pointing the compiler at the \code{include/} directory. This directory stores all necessary code, although there there are unit tests and examples provided as well.


\section{State-Space Models} \label{sec:models}

\subsection*{A Mathematical Description}\label{sec:math_desc}

\begin{figure}[ht] 
\centering
\tikzset{rectangle state/.style={draw,rectangle}} 
\begin{tikzpicture}[scale=0.2]
  \node[state, minimum size=1.5cm] (y0) {$y_{0}$};
  \node[rectangle state,minimum size=1.5cm] (x0) [below= of y0] {$x_{0}$};
  \node[state, minimum size=1.5cm] (y1) [right= of y0] {$y_{1}$};
  \node[rectangle state, minimum size=1.5cm] (x1) [right= of x0] {$x_{1}$};
  \node[state, draw=none, minimum size=1.5cm] (ydots) [right= of y1] {$\cdots$};
  \node[state, draw=none, minimum size=1.5cm] (xdots) [right= of x1] {$\cdots$};
  \node[state, minimum size=1.5cm] (yTm1) [right= of ydots] {$y_{T-1}$};
  \node[rectangle state, minimum size=1.5cm] (xTm1) [below= of yTm1] {$x_{T-1}$};
  \node[state, minimum size=1.5cm] (yT) [right= of yTm1] {$y_{T}$};
  \node[rectangle state, minimum size=1.5cm] (xT) [below= of yT] {$x_{T}$};
  \draw[->]
    (x0) edge[left] (y0)
    (x1) edge[left] (y1)
    (xTm1) edge[left] (yTm1)
    (xT) edge[left] (yT)
    (x0) edge[left] (x1)
    (x1) edge[left] (xdots)
    (xdots) edge[left] (xTm1)
    (xTm1) edge[left] (xT);
\end{tikzpicture}
\caption{diagram of a state-space model}
\label{fig:ssm_diagram}
\end{figure}
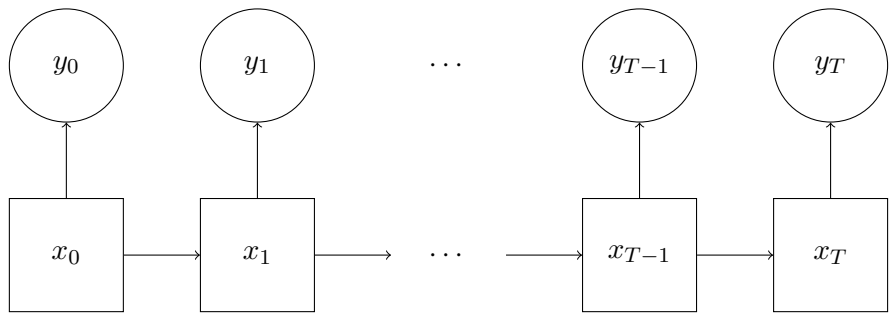

A state-space model (SSM), also known as a hidden Markov model or partially-observed Markov process, is defined by two things: an observable time series $\{y_t\}_{t=0}^T \coloneqq y_{0:T}$ of length $T +1$, and an unobservable or latent time series $\{x_t\}_{t=0}^T \coloneqq x_{0:T}$. Specifying a SSM requires the selection of several probability distributions all parameterized by a vector $\theta$. What are chosen are $\mu(x_0 \mid \theta)$, the first time's state distribution, along with the state transition densities, $\{f(x_t \mid x_{t-1},\theta)\}_{t=1}^T$, and the observation densities, $\{g(y_{t} \mid x_{t},\theta)\}_{t=0}^T$. The generative graph of this process is displayed in Figure~\ref{fig:ssm_diagram}

Constructing the joint density of the complete data, which includes both the observed and unobserved portions, is straightforward because $y_{0:T}$ are conditionally independent given the states, and because the states are assumed to possess the Markov property. Specifically, this means
\begin{eqnarray}
p(y_{0:T} \mid x_{0:T},\theta ) & = & \prod_{t=0}^T g(y_t \mid x_t,\theta ) \label{obs}
\end{eqnarray}
and
\begin{eqnarray}
p(x_{0:T} \mid \theta) & = & \mu(x_0 \mid \theta) \prod_{t=1}^T f(x_t \mid x_{t-1},\theta). \label{trans}
\end{eqnarray}
The product of (\ref{obs}) and (\ref{trans}) yields the complete data likelihood: $p(x_{0:T},y_{0:T} \mid \theta)$, and integrating out the unobserved $x_{0:T}$ from this yields the marginal likelihood $p(y_{0:T} \mid \theta)$. For more information see the tutorial by \cite{tutorial}. For a more detailed book-length description, see \cite{10.5555/1088883}, which makes use of concise measure-theoretic notation and provides many accessible proofs to its theorems.

\subsection*{Algorithmic Descriptions}\label{sec:algo_desc}

There are primarily two types of quantities of interest for users of particle filters. The first is a conditional expectation involving integration (or summation) with respect to the distribution of states conditioning on some observed data: $p(x_{0:t} \mid y_{0:t},\theta)$. Generally speaking, this task becomes computationally difficult when this distribution is not tractable. This does not always happen, it depends on how the conditional distributions defining the model were chosen by the data analyst. When it does not happen, Kalman filters and smoothers, and the assorted algorithms for discrete state-space hidden Markov models are quite prevalent.

This first type of quantity of interest can be further subdivided into "smoothing" and "filtering." Filtering is suitable for many real-time applications such as tracking and prediction, and is used to approximate expectations with respect to the distribution of the most recent state, conditioning on all information up to that time point. Smoothing, on the other hand, is used for retrospective analyses. \pkg{PF} deals exclusively with the former.

The second type of quantity consists of likelihoods. At each iteration, particle filters involve a calculation used to approximate the conditional distribution of the most recent data point, given all the previous data points. For $t \ge 1$, this can be written as $\hat{p}(y_t \mid y_{0:t-1}, \theta)$, and for $t=0$, this can be written as $\hat{p}(y_0 \mid \theta)$. Factoring the marginal likelihood as a product motivates an estimator for the entire marginal likelihood:
\begin{equation} \label{eq:1}
\hat{p}(y_{0:T}) = \hat{p}(y_0 \mid \theta) \prod_{i=1}^T \hat{p}(y_t \mid y_{0:t-1}, \theta).
\end{equation}
This quantity is often unbiased \cite{delmoral2004}, and can be useful within algorithms that perform parameter inference \cite{estimationreview}. In particular, particle Markov chain Monte Carlo \cite{pmcmc} can be programatically identical to a standard Metropolis-Hastings algorithm \cite{Metropolis1953}, \cite{10.2307/2334940} with the exception that evaluations of the likelihood are replaced by the approximations provided by Equation \ref{eq:1}.

\begin{algorithm}
\caption{Sequential Importance Sampling with Resampling (SISR)}\label{alg:SISR}
{\fontsize{7}{2}\selectfont
\begin{algorithmic}[0]
\Procedure{sisr}{$g_t,f_t,q_t,\mu$}
  \If{$t$ equals $0$}
    \For{$i=0,\ldots, N-1$}
      \State samples[0,i] $\gets$ $X_0^i \sim q_0(x_0 \mid y_0, \theta)$
      \State uNrmWts[i] $\gets \frac{ g(y_0 \mid X^i_0, \theta) \mu(X_0^i \mid \theta)}{ q_0(X_0^i\mid y_0, \theta)}$
    \EndFor
    \State lastLogCondLike $\gets \log\left[\hat{p}(y_0)\right]$
    \If{resampling criterion true}
      \State resample(samples, uNrmWts)\Comment{sets weights to uniform}
    \EndIf
  \Else\Comment{$t > 0$}
  \For{$i=0,\ldots N-1$}
    \State samples[t,i] $\gets$ $X_t^i \sim q_t(x_t \mid x_{t-1},y_t,\theta)$
    \State uNrmWts[i] $*= \frac{g(y_t\mid X_t^i, \theta)f(X_t^i\mid x_{t-1}^i,\theta)}{q_t(X_t^i\mid y_t,x_{t-1}^i,\theta)}$
  \EndFor
  \State lastLogCondLike $\gets \log\left[\hat{p}(y_t\mid y_{1:t-1},\theta)\right]$
    \If{resampling criterion satisfied}
      \State resample(samples, uNrmWts)\Comment{sets weights to uniform}
    \EndIf
  \EndIf
\EndProcedure
\end{algorithmic}
}
\end{algorithm}

\section*{ Using \pkg{PF} }

\subsection*{Getting started}

This section provides a fully worked example. It starts by describing how to download the software. The software can be obtained by cloning the Github repository here: \url{https://github.com/tbrown122387/pf}. Familiarity with \code{git} is not required. It suffices to click the green "Clone or download" button, and then click "Download ZIP."

After extracting, all the code you need to build a working program is located in the \code{include/} directory. This library is header-only, so nothing here needs to be compiled separately from your program. The only compiler requirement is that you will need to enable \code{C++11} (e.g., by including the \code{-std=C++11} flag). Second, this code depends on the \pkg{Eigen} library, and, to a lesser extent, the \pkg{Boost} library as well. 

\subsection*{Implementing a model}

Take as an example a simple stochastic volatility model with three parameters: $\beta, \phi$ and $\sigma$. These are popular for modeling a time series of the rates of return of financial instruments when the distributions of these returns have a time-varying scale. The discovery of these models is often credited to \cite{taylor82}, and they are still extensively studied. At the time of writing this, searching "stochastic volatility" on Google scholar returns 645,000 results.

For this model, we will assume the observable rate of return $y_t$ is normally distributed, but only after conditioning on the contemporaneous state random variable $x_t$. The mean parameter of this normal distribution will remain fixed at $0$; however, the scale of this distribution will vary with the evolving $x_t$. When $x_t$ is relatively high, the returns will have a high conditional variance and be "volatile." When $x_t$ is low, the returns will be much less volatile.

We write this observational equation as
\begin{eqnarray}
y_t = \beta e^{x_t/2} z_t
\end{eqnarray}
where the collection $\{z_t\}_{t=0}^T$ are independent and identically distributed normal random variates.

The state evolves randomly through time as a simple autoregressive process:
\begin{eqnarray}
x_t = \phi x_{t-1} + \sigma z'_t.
\end{eqnarray}
The collection $\{z'_t\}_{t=1}^T$ are also assumed to be independent and identically distributed normal random variates. At time $0$, when there is no previous state value, we assume the first state follows a mean zero normal distribution with $\sigma^2/(1-\phi^2)$, and for simplicity, all of our proposal distributions match the state's distribution at the time.

The file \code{examples/svol_sisr.h} provides a fully worked example of writing a class template called \code{svol_sisr} for this model-algorithm pair. This is a class template, as opposed to a class, so that means the decision of what to pass in as template parameters has been pushed back to the code that instantiates and uses these particle filters. Even though now this code is messier than a plain class, this will benefit the data analyst in the long run because there will likely only be one file to change. Last, the SISR algorithm's header is included with \verb|#include "sisr_filter.h"|.

\begin{leftbar}
Our choice of proposal distribution means that this algorithm satisfies the definition of the \emph{Bootstrap Filter}. A demonstration of how to use a specially designed class template is given in \code{examples/svol_bs.h}.
\end{leftbar}

\begin{Code}
template<size_t nparts, size_t dimx, size_t dimy, typename resampT, typename float_t>
class svol_sisr : public SISRFilter<nparts, dimx, dimy, resampT,float_t>
{
public:
    using ssv = Eigen::Matrix<float_t, dimx, 1>;
    using osv = Eigen::Matrix<float_t, dimy, 1>;

    float_t m_phi;
    float_t m_beta;
    float_t m_sigma;

    rvsamp::UnivNormSampler<float_t> m_stdNormSampler;

    svol_sisr(const float_t &phi, const float_t &beta, const float_t &sigma);

    float_t logMuEv (const ssv &x1);
    ssv q1Samp (const osv &y1);
    float_t logQ1Ev (const ssv &x1, const osv &y1 );
    float_t logGEv (const osv &yt, const ssv &xt );
    float_t logFEv (const ssv &xt, const ssv &xtm1 );
    ssv qSamp (const ssv &xtm1, const osv &yt );
    float_t logQEv (const ssv &xt, const ssv &xtm1, const osv &yt );
};
\end{Code}

After the type aliases, the first thing to notice inside the body of the class template is that the parameters are simple data members of the model. These need to be data members stored inside of the class because these numbers will be referenced in the class's methods' definitions. Typically, the user will choose the template parameter \code{float_t} to be either type \code{double} or type \code{float}.

Second is the declaration of the member \code{m_stdNormSampler}. This will be used in our functions as well, but only in the ones that sample random variates. All of the distributions we sample from are normal distributions; if that wasn't the case, it would be easy to add more members that sample from different distributions. This particular object is of type \code{rvsamp::UnivNormSampler<float_t>}. The class template we are making use of, \code{rvsamp::UnivNormSampler}, is defined in the file \code{include/rv_samp.h}.

Third, there are function signatures. One is the constructor, which is simple to implement, and the rest are the statistical distributions that define the model. The names of the mathematical functions match the mathematical notation that was introduced in Section~\ref{sec:math_desc} and Section~\ref{sec:algo_desc}.

To define the constructor for the class template, be sure to 1.) call the base class's constructor, and 2.) allow user-provided model parameters to be stored as data members.
\begin{Code}
template<size_t nparts, size_t dimx, size_t dimy, typename resampT, typename float_t>
svol_sisr<nparts,dimx,dimy,resampT,float_t>::svol_sisr(const float_t &phi,
                                                       const float_t &beta,
                                                       const float_t &sigma)
    : SISRFilter<nparts, dimx, dimy, resampT,float_t>()
    , m_phi(phi), m_beta(beta), m_sigma(sigma) { }
\end{Code}
Note that \code{resampT}, the chosen resampling method, is provided as a type. That type will be used to declare a data member in the instantiation of the base class template.

Implementations of the class template methods are also in \code{examples/svol_sisr.h}. The methods that {\bf evaluate} random variables make use of function templates provided in \code{include/rv_eval.h}, and methods that sample pseudo-random numbers make use of class templates provided in \code{include/rv_samp.h}.

One reason the example keeps the specific model's particle filter as a class template is to demonstrate the use of \emph{trailing return types}, which is a feature introduced in \code{C++11} \cite{ISO:2012:III}. Take for instance the definition of the proposal distribution $q(x_0 \mid y_0, \theta)$:
\begin{Code}
template<size_t nparts, size_t dimx, size_t dimy, typename resampT, typename float_t>
auto svol_sisr<nparts,dimx,dimy,resampT,float_t>::q1Samp(const osv &y1) -> ssv
{
    ssv x1samp;
    x1samp(0) = m_stdNormSampler.sample() * m_sigma / std::sqrt(1.-m_phi*m_phi);
    return x1samp;
}
\end{Code}
This is necessary because the return type of this method, \code{ssv}, will depend on one of the class template's template parameters (i.e., \code{dimx}).

\subsection*{Using a model class template}

Instantiating an object with the class template defined in \code{examples/svol_sisr.h} is even easier. Just provide the template parameters and the constructor parameters in the correct order. For example,
\begin{Code}
svol_sisr<5000,1,1,mn_resampler<5000,1,double>,double> sisrsvol(.91,.5,1.0);
\end{Code}
instantiates the object called \code{sisrsvol}, which performs the SISR algorithm for the stochastic volatility model using multinomial sampling on $5,000$ particles. It sets $\phi = .95$, $\beta = .5$ and $\sigma = 1$.

This example is just for pedagogical purpose. After seeing that these type names can be quite long, it is good practice to use typedefs or type aliases, and to avoid hard coding parameter choices as much as possible. A fully-worked example of implementing our stochastic volatility model with a SISR algorithm, along with the same model using different particle filters, is provided in \code{examples/svol_comparison.cpp}, and is discussed later in section~\ref{sec:illustrations}.

\subsection*{Advantages and disadvantages of \proglang{C++} templates}

The data analyst must choose a model, a type of particle filter, a type of resampling, the number of particles, and the aforementioned proposal distributions. All of this information the user will supply when he writes the header file for his model's class. In the provided examples, our particle filter subclasses a class template instantiation of \code{SISRFilter}. This means that the number of particles, the dimension of the state vector, the dimension of the observations, the type of resampling, and the type of floating point numbers to use inside all of the vectors and matrices are all known at {\bf compile time}, which has a number of advantages.

First, if a data analyst knows how many particles she will end up sampling at every iteration, that means she can use the \code{std::array} container. These containers are possibly faster than \code{std::vector}-based containers, although this will depend on the number of particles, and the compiler flags used. They also allow for data analyst to choose whether the particle filter object is to be stored on the stack or on the heap. Finally, they prohibit the costly and unnecessary resizing of the container.

The template parameters for the dimension of the observed vectors, the unobserved vectors, and the type of floating point scalars are used as the first three template arguments to the \code{Eigen::Matrix} class template \cite{eigenweb}. Again, because they are known at compile time, these sizes can be used to opt for fixed size vectors instead of "dynamic" ones. This has the benefit of speed in many situations, particularly for smaller sizes. Moreover, manipulating single precision floating point numbers is computationally cheaper than manipulating double precision ones. This savings can add up if there are many particles being used. There are downsides to this, though. If the particle filter objects are being stored on the stack, there is a possibility of stack overflow errors.

\pkg{PF} offers several particle filters to subclass (for more information, see Table~\ref{tab:overview}). In addition to the three algorithms used in the example, there are class templates for closed-form filtering algorithms such as Kalman filtering \cite{kalman}, \cite{shumwaystoffer} and the HMM filter \cite{baum1966}, \cite{rabiner}, as well as four different types of "Rao-Blackwellized" or "marginalized filters" \cite{rbpfandrieudoucet}, \cite{Chen00mixturekalman}.

\begin{table}[t!]
\centering
\begin{tabular}{llll}
\hline
 Name     & Sampling or Closed-Form &Available resamplers       &  \\ 
\hline
\verb|BSFilter|                  & sampling-based                 &multinomial, residual, stratified, systematic &  \\
\verb|SISRFilter|                & sampling-based                 & multinomial, residual, stratified, systematic &  \\
\verb|APF|                       & sampling-based                 & multinomial, residual, stratified, systematic &  \\
\verb|kalman|                       & closed-form                    & multinomial, residual, stratified, systematic &  \\
\verb|hmm|                                    & closed-form                    & multinomial, residual, stratified, systematic &  \\
\verb|rbpf_kalman_bs| & both                           & multinomial                                   &  \\
\verb|rbpf_hmm_bs|    & both                           & multinomial                                   &  \\
\verb|rbpf_kalman|     & both                           & multinomial                                   &  \\
\verb|rbpf_hmm|        & both                           & multinomial                                   & \\
\hline
\end{tabular}
\caption{\label{tab:overview} Overview of the available particle filtering algorithms.}
\end{table}

The primary difference between them is which proposal distribution the user decides to use to mutate old samples into new ones. At time $0$, this distribution is denoted by $q_0(x_0 \mid y_0, \theta)$, and for $t \ge 1$, it is denoted by $q_t(x_t \mid x_{t-1}, y_t, \theta)$. It is best to start with a discussion of the sequential importance sampling with resampling (SISR) algorithm , because the bootstrap filter \cite{gordonetal} (also known as the condensation algorithm) \cite{blake1997condensation} and the auxiliary particle filter \cite{auxpf} are both special cases of it. The pseudo-code for SISR is given in Algorithm~\ref{alg:SISR}.



\section{Illustrations} \label{sec:illustrations}

This section will follow up on the example provided in the \code{examples/} directory, describing some of the higher level details of the program.

As it is in every \proglang{C++} program, the entry point for the example program is \code{main.cpp}. When a user provides a $1$ from the command line, \code{main()} calls \code{run_svol_comparison()} is called, which is a function declared in \code{examples/svol_comparison.h}, and defined in \code{examples/svol_comparison.cpp}. Its implementation is described now. The file in its entirety can be found in \proglang{PF}'s Github repository.

First, comes all of the include statements. We include this file's header file, a header that provides a function to read in data, a header for each of the three different implementations of each particle filter-model pair, and a header file for our resampler types.

\begin{Code}
#include "svol_comparison.h"
#include "data_reader.h"
#include "svol_bs.h"
#include "svol_apf.h"
#include "svol_sisr.h"
#include "resamplers.h"
\end{Code}

Second, we \verb|#define| the template parameters of our particle filters. We do this at the top of the file, so that they can be changed quickly and easily later on when the user wants to change the particle filter options.

\begin{Code}
#define dimstate 1
#define dimobs   1
#define numparts 5000
#define FLOATTYPE float
\end{Code}

Third come some type aliases. Their names are abbreviations that stand for "state size vector", "observation sized vector" and "matrix," respectively. 

\begin{Code}
void run_svol_comparison(const std::string &csv)
{
    using ssv = Eigen::Matrix<FLOATTYPE,dimstate,1>;
    using osv = Eigen::Matrix<FLOATTYPE,dimobs,1>;
    using Mat = Eigen::Matrix<FLOATTYPE,Eigen::Dynamic,Eigen::Dynamic>;
\end{Code}

Fourth, we instantiate the parameters of the model. These will be shared by all three particle filters.

\begin{Code}
    FLOATTYPE phi = .91;
    FLOATTYPE beta = .5;
    FLOATTYPE sigma = 1.0;
\end{Code}

Fifth, we instantiate a stochastic volatility model's bootstrap filter, auxiliary particle filter, and SISR filter. All of them use multinomial resampling.

\begin{Code}
    svol_bs<numparts, dimstate, dimobs,
            mn_resampler<numparts,dimstate,FLOATTYPE>,
            FLOATTYPE> bssvol(phi, beta, sigma);
    svol_apf<numparts,dimstate,dimobs,
             mn_resampler<numparts,dimstate,FLOATTYPE>,
             FLOATTYPE> apfsvol(phi,beta,sigma);
    svol_sisr<numparts,dimstate,dimobs,
             mn_resampler<numparts,dimstate,FLOATTYPE>,
             FLOATTYPE> sisrsvol(phi,beta,sigma);
\end{Code}

Finally, some time series data is read in, and stored in a \code{std::vector} of \code{Eigen} vectors. The \code{std::vector} of lambda functions is provided as a second argument to each particle filter's \code{filter} method. The provided lambda function is called on each particle/sample, and at each time point, the weighted average of all these numbers is returned as an approximation to the filtering expectation $E[x_t \mid y_{0:t}, \theta].$ The for loop iterates over the data, and at each time point, prints out each model's estimate of the scalar-valued quantity would correspond with a real-time estimate of the current volatility level for this financial instrument, as well as an estimate for the logarithm of the conditional likelihood $p(y_t \mid y_{0:t-1}, \theta)$.

\begin{Code}
    std::vector<osv> data = readInData<FLOATTYPE,dimobs>(csv);
    auto idtyLambda = [](const ssv& xt) -> const Mat
    {
        return xt;
    };
    std::vector<std::function<const Mat(const ssv&)>> v;
    v.push_back(idtyLambda);
    for(size_t row = 0; row < data.size(); ++row){
        bssvol.filter(data[row], v);
        apfsvol.filter(data[row], v);
        sisrsvol.filter(data[row], v);

        std::cout << bssvol.getExpectations()[0] << ", "
                  << bssvol.getLogCondLike() << ", "
                  << apfsvol.getExpectations()[0] << ", "
                  << apfsvol.getLogCondLike() << ", "
                  << sisrsvol.getExpectations()[0] << ", "
                  << sisrsvol.getLogCondLike() << "\n";
    }
}
\end{Code}

The above code is built and run with the following commands. It writes all output to a file named \code{~/Desktop/sample_output.txt}. Here we are using the utility \proglang{make}. This code was run on the Bourne Again Shell in Ubuntu 18.0.4, but it will likely work on other Mac and Linux machines. If it does not, there might be some small adjustments that need to be made to the file \code{examples/Makefile}--for example, this file assumes the \code{Eigen} headers are located at \code{/usr/include/eigen3}.
\begin{Code}
cd ~/pf/examples/
make clean && make
./examples 1 data/svol_y_data.csv > ~/Desktop/sample_output.txt
\end{Code}

\begin{figure}[t!]
\centering
\includegraphics{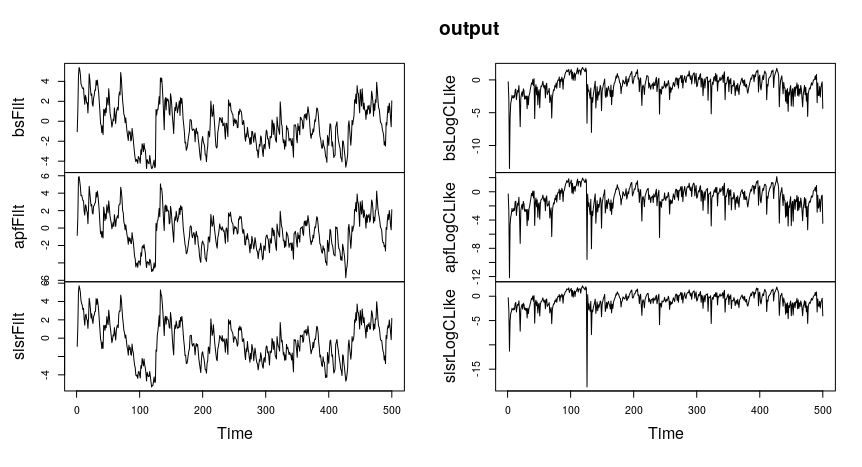}
\caption{\label{fig:output} Comparison of each model's estimates of the filter means, $E[x_t \mid y_{0:t}, \theta]$, and the conditional log-likelihoods, $\hat{p}(y_t \mid y_{0:t}, \theta)$.}
\end{figure}

The output can then be more closely evaluated. For example, the following \proglang{R} code would produce a comparison of the three particle filter's outputs:
\begin{Code}
output <- read.csv("~/Desktop/sample_output.txt", header=F)
colnames(output) <- c("bsFilt", "apfFilt", "sisrFilt",
                 "bsLogCLike", "apfLogCLike", "sisrLogCLike")
plot.ts(output, plot.type = "multiple", nc = 2)
\end{Code}
This plot is displayed in Figure~\ref{fig:output}.


\section{Summary and discussion} \label{sec:summary}

This article has outlined the design of, and provided a fully-worked example for, the \pkg{PF} library, a \proglang{C++} header-only template library for particle filtering. This library will continue to be developed into the foreseeable future, so feature requests and bug reports can be provided after visiting \href{https://github.com/tbrown122387/pf/issues}{https://github.com/tbrown122387/pf/issues}.


\section*{Computational details}

The results in this paper were obtained with the \pkg{PF}~1.0.1 library, using the \code{gcc 7.4.0} compiler on the \code{Ubuntu 18.04} operating system. Compilation of the \code{examples} program was also successfully completed with the \code{clang++-6.0} compiler, as well.


\bibliography{refs}


\end{document}